\documentclass{aa}    
\usepackage{graphics} 
 
\begin{document} 
 
\thesaurus{ 3(08.06.2; 
		11.09.1 The Cartwheel; 
		11.09.2; 
		11.19.3; 
		13.09.3)  
             } 
\title{Dust in the Wheel: The Cartwheel Galaxy in the 
Mid-IR\thanks{Based on observations with
ISO, an ESA project with instruments funded by ESA Member States
(especially the PI countries: France, Germany, the Netherlands and the
United Kingdom) with the participation of ISAS and NASA.}
} 
   
\author{V. Charmandaris 
	\inst{1,2},  
	O. Laurent  
	\inst{2}, 
	I.F. Mirabel  
	\inst{2}, 
	P. Gallais  
	\inst{2}, 
	M.~Sauvage 
	\inst{2}, 
	L.~Vigroux  
	\inst{2}, 
	C.~Cesarsky 
	\inst{2} 
	\and 
	P.N. Appleton 
	\inst{3} 
	} 
\offprints{V.Charmandaris, v.charmandaris@obspm.fr} 
  
\institute{ 
	Observatoire de Paris, DEMIRM, 61 Av. de l'Observatoire, 
	F-75014, Paris, France 
	\and 
	Service d'Astrophysique, CEA-Saclay, F-91191, Gif sur 
	Yvette Cedex, France 
	\and 
	Department of Physics \& Astronomy, Iowa State University, 
	Ames IA, 50011, U.S.A. 
	} 
\date{Received 9 July 1998 / Accepted 13 October 1998} 
\authorrunning{Charmandaris et al.} 
\titlerunning{The Cartwheel Galaxy in the Mid-IR}
\maketitle 
 
\begin{abstract}   
We present mid-infrared images at $6.7 \mu m$ and $15 \mu m$ of ``The
Cartwheel'' (AM~0035-33), the prototypical collisional ring
galaxy. The observations, taken with ISOCAM, reveal the distribution
of hot dust in the galaxy and its two companions in the
north-east. The intensity of the mid-IR emission from the outer star
forming ring of the Cartwheel shows considerable azimuthal variation
and peaks at the most active H$\alpha$ region of the ring. The
15\,$\mu m$ to 6.7\,$\mu m$ flux ratio of 5.2 is the highest among all
the galaxies of our sample. A surprising result of our observations is
the discovery of significant emission from the inner regions of the
galaxy, including the inner ring, spokes and nucleus, where previously
only low level H$\alpha$ emission had been reported. At 6.7\,$\mu m$,
this emission is stronger than the one from the outer star forming
ring, and at 15\,$\mu m$, it represents 40\% of the emission from the
outer ring. The H$\alpha$ to mid-IR flux ratios from the inner regions
are consistent with the heating of grains from weak star formation
activity.

\end{abstract} 
  
\section{Introduction} 
 
Collisional ring galaxies, of which the Cartwheel is the
``prototypical'' candidate, are believed to form when an ``intruder''
galaxy passes through the center of a rotating disk of a larger
``target'' galaxy (Lynds \& Toomre \cite{lt}; Theys \& Spiegel
\cite{ts}; Appleton \& Struck-Marcell \cite{rings} and references
therein). The perturbation triggers a radially expanding ring-like
density wave on the disk, causing massive star formation in the ring
(see Appleton \& Marston \cite{ring97}). The symmetry and well defined
dynamical history of ring galaxies has made them ideal candidates for
studies of the phase transition of the interstellar medium due to
collisionally induced star formation.
   
The Cartwheel galaxy was discovered by Zwicky (\cite{zwicky}) at a
distance of 121 Mpc (H$_{o}$=75\,km\,s$^{-1}$\,Mpc$^{-1}$). It has a
bright outer ring and an inner ring which is connected to the outer
one with a series of spokes (see Higdon \cite{jim_hi} Fig.~1). Three
small companion galaxies located north and north-east of the ring
complete the group.  It is still unclear which of the companions is
the culprit for the creation of the Cartwheel ring (Davies \& Morton
\cite{dm}; Struck-Marcel \& Higdon \cite{curt_jim}), but it is likely
that each of them contains sufficient mass to trigger the generation
of the star forming ring (see discussion by Appleton \& Struck-Marcell
\cite{rings}). On-going efforts to simulate the dynamics of the system
have been focused on the nearby companion G2 (Bosma priv. comm.) and
on the most distant one, G3, (Struck \cite{curt}; Horellou priv. comm.)
which seems to be connected to the Cartwheel with a plume of
\ion{H}{i} gas (Higdon \cite{jim_hi}).
 
Due to its unique morphology, the Cartwheel has been the subject of
early optical studies (Theys \& Spiegel \cite{ts}; Fosbury
\& Hawarden \cite{fh}) as well as dynamical modeling (Struck-Marcel \& 
Higdon \cite{curt_jim}; Hernquist \& Weil \cite{hw}). The outer ring of 
the Cartwheel ($\sim$ 70$\arcsec$ in diameter) is expanding, has blue 
colours and is populated by massive star-forming regions (Higdon 
\cite{jim_ha}, Amram et al. \cite{amram}).  Most of the star 
formation though, appears to occur in a localized area of a few 
\ion{H}{ii} complexes in the southern sector of the ring. The 
H${\alpha}$ emission, as well as the 20cm and 6cm radio continuum
emission vary as a function of the azimuth along the ring and peak in
the same region of the southern sector (Higdon \cite{jim_hi}).
Optical and near-IR imaging show strong radial colour gradients in the
disk behind the outer ring, which may trace the evolution of the
stellar population in the wake of the density wave (Marcum et
al. \cite{marcum} ). Broad band images obtained with the HST (Borne et
al. \cite{borne}; Appleton \cite{iau}) reveal in unprecedent
detail the distribution of massive young clusters around the outer
ring, as well as the diffuse and knotty structure of the so called
``spokes''.
 
The inner ring ($\sim$ 18$\arcsec$ in diameter) and nucleus of the
Cartwheel seem gas-poor, have very little \ion{H}{i} (Higdon
\cite{jim_hi}), and no CO emission has been detected so far (Horellou
et al. \cite{horellou}).  The failure to detect molecules from the
galaxy may be a matter of sensitivity, or low metalicity since recent
H$\alpha$ observations (Amram et al. \cite{amram}) indicate that there
is some low-level star formation in the central regions. Furthermore,
HST observations clearly show the presence of well defined dust lanes,
as well as some interesting compact blue comet-like regions on and
near the inner ring (Struck et al. \cite{struck}). The heads of the
cometary structures were suggested to be regions of star formation
triggered by the passage of dense clouds moving supersonically through
the inner-ring gas. Such clouds may result from infall from the \ion{H}{i}
plume, or cloud-cloud collisions in the disk.

Although the galaxy as a whole was detected by IRAS at longer
wavelengths (Appleton \& Struck-Marcell \cite{irasasm}), our ISOCAM
mid-IR observations represent the first detection and mapping of the
distribution of the hot dust throughout the galaxy and its nearby
companions.
 
  
\section{Observations and Data Reduction}   
 
The Cartwheel group was observed with ISOCAM (Cesarsky et
al. \cite{cesarsky}) on November 23 1996 (ISO revolution 372).  It was
part of the ISO (Kessler et al. \cite{kessler}) guaranteed time
program CAMACTIV (PI. I.F. Mirabel) which prime goal was the mid-IR
imaging of more than 20 nearby active/interacting galaxies.  Two broad
band filters centered at 6.75\,$\mu m$ (LW2), and 15\,$\mu m$ (LW3), with
a lens resulting in a 6$\arcsec$ pixel field of view, were used to
create a 2$\times$2 raster map. The mid-IR maps cover the Cartwheel
galaxy and the two nearby companions G1 and G2. More details on the
observational parameters are presented in Table~\ref{param}.
 
\begin{table}[h]  
\caption[ ]{Observational  parameters} 
\label{param} 
\begin{flushleft}  
\begin{tabular}{lll}  \hline 
Name: 		& The Cartwheel \\ 
Coordinates:	& a(J2000)		& 00$^h$ 37$^m$ 41.62$^s$\\ 
		& $\delta$(J2000)	&-33$\degr$43$\arcmin$00.4$\arcsec$ \\ 
Observations:	& Instrument		& ISOCAM\\ 
		& Date			& November 23, 1996\\ 
		& Field of view		& 4.2$\times$ 4.2 arcmin\\ 
		& Pixel size		& 6$\arcsec$\\ 
Filters:	& 6.75$\mu m$ (LW2)	& range: [5 - 8.5] $\mu m$\\ 
		& 15$\mu m$ (LW3)	& range: [12 - 18] $\mu m$\\ 
Exposure times:	& Number of frames	& 141 per filter\\ 
		& Frame exposure time	& 2.1 s\\ 
		& Total LW2 exposure	& 5\,min \\ 
		& Total LW3 exposure	& 5\,min \\ 
 
\hline    
\end{tabular} 
\end{flushleft}   
\end{table} 
 
The standard data reduction procedures described in the
ISOCAM\footnote{The ISOCAM data presented in this paper were analyzed
using ``CIA'', a joint development by the ESA Astrophysics Division
and the ISOCAM Consortium led by the ISOCAM PI, C. Cesarsky, Direction
de Sciences de la Mati\`ere, C.E.A., France.}  manual were followed
(Delaney \cite{isoman}). Dark subtraction was performed using a model
of the secular evolution of ISOCAM's dark current (Biviano et
al. \cite{biviano}). Cosmic rays were removed using a multi-resolution
median filtering method (Starck et al. \cite{starck}) while the memory
effects of the detector were corrected using the so-called IAS
transient correction algorithm which is based on an inversion method
(Abergel et al. \cite{abergel}). The final raster was constructed
after using the instrumental flat fields and correcting for the lens
field distortion. These methods and their consequences are discussed
in detail in Starck et al. (\cite{isocam}).
 
The ISOCAM LW2 filter mainly samples mid-IR flux originating from the
Unidentified Infrared Bands (UIBs), centered at 6.2, and 7.7\,$\mu
m$. This feature emission is attributed to stretching modes of
2-dimensional molecules (often called polycyclic aromatic hydrocarbons,
or PAHs) having C--C and C--H bonds. It may also contain some
contribution from the long wavelength blackbody tail of the stellar
photospheric emission. The LW3 filter though, is almost two times
wider than the LW2 and it is principally sensitive to the presence of
the thermal continuum emission of very small grains. Regions of
massive young stars can heat very efficiently the surrounding dust
grains and create a strong thermal mid-IR continuum (i.e. Vigroux et
al. \cite{chef}). This continuum appears at 12\,$\mu m$, its slope
steepens as the intensity of the ionizing field increases, and it
dominates the mid-IR emission up to 18\,$\mu m$ (the long wavelength
detection limit of ISOCAM). Several forbidden lines and few UIBs may
also appear in the wavelength range covered by LW3, but their
contribution to the total flux is usually negligible when the thermal
continuum is present.
  
\section{The Morphology of the Mid-IR Emission} 
 
\begin{figure*} 
\resizebox{18cm}{!}{\includegraphics{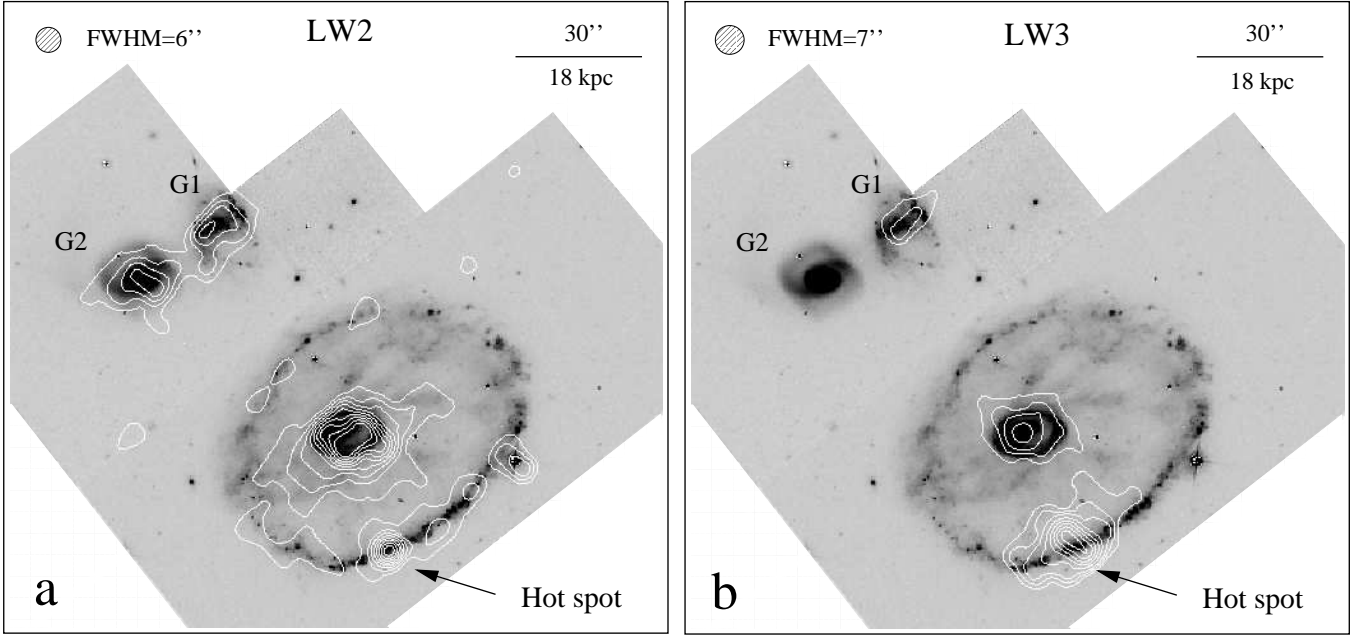}}  
 \caption{
a) Contour map of the ISOCAM LW2 emission overlaied on an HST
wide I band image. The contour levels are from 0.1 to 0.5 mJy/pixel with a
step of 0.05 mJy/pixel.  b) Contour map of the ISOCAM LW3 emission
overlaied on the same HST image. The contour levels are
0.2,0.3,0.4,0.6,0.8,1,1.2 and 1.4 mJy/pixel. North is up and East is 
left in both images.
}
  \label{lw23} 
\end{figure*}

As one can observe from Fig. \ref{lw23}, despite of not being detected
in the IRAS 12$\mu m$ and 25$\mu m$ bands (upper limits of $\sim$0.1
Jy in both bands), there is mid-IR emission from the Cartwheel group
(see Table~\ref{phot}) and it displays an interesting spatial
distribution.  In the following subsections, we discuss the morphology
and nature of the mid-IR emission from the two companions, the outer
star forming ring of the Cartwheel, as well as from the central
regions of the ring galaxy.

\subsection{The Companions} 
 
The morphological types of the two nearby companions of the Cartwheel
are distinctly different. The eastern companion (G2) is an early type
S0 galaxy.  No \ion{H}{i} is directly associated with this galaxy
(Higdon \cite{jim_hi}) and no star formation activity has been
detected (Amram et al. \cite{amram}).  On the contrary, the western
companion (G1), a late type galaxy with somewhat irregular spiral
arms, has 2.7x10$^{9}$ M$_{\odot}$ of \ion{H}{i} and emits strongly in
H$\alpha$, which indicates the presence of a young stellar population
(see Fig.1 of Amram et al. \cite{amram}). The LW2 fluxes of the two
galaxies, presented in Table~\ref{phot}, are weak but very
similar. This is not true though for the LW3 emission where only the
western companion (G1), is detected.
 
This contrast on the LW2 over LW3 emission from the two companion 
galaxies is in agreement with our current understanding of the typical 
mid-IR signature in galaxies (Vigroux \cite{vigroux}). In early type 
galaxies the mid-IR emission is dominated by the old stellar 
population. The LW2 flux is typically 6 times stronger than the LW3 
emission which is consistent with temperatures of $\sim$5000~K (Madden 
\cite{madden}). Since the LW2 emission of G2 is just $\sim$4 times 
stronger than the rms noise of the LW3 emission any expected stellar
contribution to the LW3 filter measurement would be below our
detection limit.  The ratio of LW3 to LW2 flux of G1 is $\sim$1.4, a
value found in most normal late type galaxies.

\subsection{The Cartwheel Outer Ring}

Our ISOCAM observations reveal that one-third of the circumference of
Cartwheel's outer ring is detected in the LW2 band (see
Fig.~\ref{lw23}a). The detected segment of the ring corresponds to the
brightest part of the optical ring as defined by the H$\alpha$ image
of Higdon (\cite{jim_ha}). More than 80\% of the total H$\alpha$ flux
from the Cartwheel outer ring originates from its south-eastern
segment.  About 60\% of the LW2 flux from the outer ring comes from a
hot-spot in the ring, unresolved to our observations, which
corresponds to the position of two large \ion{H}{ii} region complexes
(Higdon \cite{jim_ha}).

Interestingly, in the LW3 filter, only the hot-spot is detected in the
outer ring (Fig.~\ref{lw23}b), and the only other emission from the
Cartwheel is from the central regions (see Table~\ref{phot} and
discussion below). The LW3/LW2 flux ratio, often used as a diagnostic
of the intensity of the radiation field, of the hot-spot is 5.2, a
value which is among the highest detected in all interacting galaxies
of the CAMACTIV sample. For comparison, the highest LW3/LW2 flux ratio
detected in the interacting galaxy NGC\,4038/39 is only 2.6 and it is
found in the region where the two disks overlap (Vigroux et
al. \cite{chef}; Mirabel et al. \cite{felix}).

The lack of detection in the  LW3 filter from regions of the ring
other than the hot-spot is intriguing. What is clear is that if the
rest of the ring had the same ratio of LW3/LW2 (i.e. 5.2) as the
hot-spot, then it would have been easily detected in the LW3
band. Therefore, using the rms noise of the LW3 image, as well as the
total LW2 flux of the ring (excluding the hot-spot -- see
Table~\ref{phot}), we set a 2$\sigma$ {\it upper limit} on the LW3 to
LW2 ratio of 1.45 in the optically bright regions away from the
hot-spot. What could be responsible for the difference in flux ratio
for the hot-spot, as compared with the other ring regions?

A clue as to why the hot-spot is different comes from a comparison
with radio and optical observations. The hot-spot coincides with the
peak of the 20cm and 6cm radio continuum emission from the ring, and,
unlike the rest of the outer ring, shows a sudden drop in \ion{H}{i}
surface density at that point (Higdon \cite{jim_hi}). Assuming that
\ion{H}{i} absorption is not responsible for the drop in \ion{H}{i}
surface density, then this suggests that most of the cool gas in the
hot-spot region is either ionized and/or has already been converted
into stars. The very powerful UV radiation field in the hot-spot
implied by its radio and H$\alpha$ emission -- the H$\alpha$ surface
density is at least a factor of 3 times higher there compared with the
other regions of the ring (Higdon \cite{jim_ha}) -- would be expected
to illuminate a {\em larger volume} of interstellar grains, thereby
boosting the thermal continuum at longer wavelengths sampled by the
LW3 filter relative to the shorter wavelength PAH-dominated LW2
emission. For an idealized spherical \ion{H}{ii} region of radius $R$,
its volume is proportional to $R^{3}$ while the volume of the
associated photo-dissociation-region (PDR) which has a shell shape
would vary as $R^{2}$.  Hence, the filling factor of the \ion{H}{ii}
regions in a given giant molecular cloud of the ISM would increase
with a higher rate than the corresponding the PDRs.  Moreover, in a
strong UV environment, the dissociation of PAH molecules would further
diminish their contribution to the LW2 flux
(i.e. Vigroux~\cite{chef}).  These factors can provide a simple
explanation for the larger-than-normal ratio of LW3/LW2 in the
hot-spot region based purely on a difference in the strength of the
local radiation field.

An explanation for the variable ratio of LW3/LW2 fluxes in the ring
may be a purely geometrical one. It is likely that the very strong
star formation in the south-western segment of the ring is sufficient
to create local galactic fountains which would lift dust grains to
large vertical disk scale-heights, the typical size of which can be
calculated (i.e. Tenorio-Tagle \& Bodenheimer (\cite{ttb}).  Based on
the 20cm radio continuum flux the type II supernova (SN) rate on the
outer ring is $\mu_{SN}=0.1$\,yr$^{-1}$ (Higdon \cite{jim_hi}). Since
the density wave is propagating with $\sim$40 km\,s$^{-1}$ across the
disk, for an instantaneous burst model the total number of supernovae
created on the outer ring of the Cartwheel ($\sim$5\,kpc in width) is
$\sim$10$^{7}$ over a period of 1.2$\times10^{8}$\,yrs. Assuming a
typical energy output per supernova of 10$^{51}$\,ergs, an ISM density
of 1 cm$^{-3}$, and that the SN are uniformly distributed on the ring,
we estimate a vertical scale up to $\sim$4\,kpc (see Tenorio-Tagle \&
Bodenheimer \cite{ttb}, section 3.1.1). The previous calculations,
though rough, since they do not consider the finite thickness of the
galaxy disk, clearly suggest that the gas/dust of the ring will be
distributed in a torus-like shape.

This ``puffing'' of the ring in combination with the stellar winds of
the forming stars would result in making the star forming regions
density bounded and diluting of the UV field in the local ISM. As
distance from the \ion{H}{ii} regions increases vertically, the
heating flux would be lowered, and the small grain emission (and as a
result the LW3 flux) would decrease exponentially. On the other hand,
since the LW2 band is dominated by thermally-spiked PAH features, its
emission strength would decrease almost linearly to the heating
intensity (Sauvage et al.~\cite{sauvage}), making LW2 stronger
relatively to LW3. This could also explain why even though more than
95\% of the detected star formation is clearly taking place in the
outer ring of the Cartwheel (Amram et al.~\cite{amram}) the mid-IR
fluxes of the outer ring and nuclear regions are comparable.

\begin{table}[h]
\caption[ ]{Mid-infrared photometry of the Cartwheel group} 
\label{phot} 
\begin{flushleft}  
\begin{tabular}{lcccc} 
\hline\\ 
Region & Area & LW2 & LW3   & $\frac{LW3}{LW2}$ \\ 
        & arcsec$^{2}$& mJy       & mJy & \\ 
\hline 
Nucleus 	& 432	& 3.6 $\pm$ 0.14   & 3.8 $\pm$ 0.38	& 1.1\\ 
Hot Spot  	& 324 	& 1.4 $\pm$ 0.12   & 7.5 $\pm$ 0.33 & 5.2\\ 
Ring (total)  	& 540	& 2.3 $\pm$ 0.16   & 8.9 $\pm$ 0.43 & (3.8)\\ 
Cartwheel   & 972   & 5.9 $\pm$ 0.21   & 12.7 $\pm$ 0.57& (2.2)    \\
G1      	& 324 	& 1.3 $\pm$ 0.12   & 1.9 $\pm$ 0.33 & 1.4\\ 
G2      	& 324 	& 1.6 $\pm$ 0.12   & $<\,0.3$  	& $<\,0.2$\\ 
\hline 
RMS/pixel & & 0.04  & 0.11 \\ 
\hline 
\end{tabular} 
\end{flushleft}  
\end{table}

\subsection{The Inner Ring and Nucleus}

Until recently, it was believed that the outer ring of the Cartwheel
was the only one that showed signs of star formation activity (Fosbury
\& Hawarden \cite{fh}; Higdon \cite{jim_ha}). However, our new observations
show conclusively that the inner ring is strongly detected with
ISOCAM. The low spatial resolution of our images does not allow us to
resolve the inner ring and dust lanes observed by HST around the
nucleus of the Cartwheel (Struck et al. \cite{struck}).  Although
centrally concentrated, the emission from the central regions is
resolved and a faint filament of LW2 emission extending to the west of
the nucleus follows one of the ``spokes'' seen in the optical images.
However, it was initially a surprise for us to detect such a strong
signal from the center of the Cartwheel, which had previously been
believed to be almost devoid of nuclear activity (Fosbury \& Hawarden
\cite{fh}, Higdon \cite{jim_ha}). The center of the Cartwheel exceeds 
the outer star-forming ring by a factor of 1.6 in the LW2 filter while
it has 0.4 times the flux of the outer ring in the LW3 filter. The
nuclear region has mid-IR colours similar to those of late type
galaxies (LW3/LW2\,=\,1.1, see Table~\ref{phot}), although optically,
it has very red colours suggestive of an older stellar population
(Struck et al. \cite{struck}).
 

We have mentioned that Amram et al. (\cite{amram}) have recently
detected low-level H$\alpha$ emission from the Cartwheel center, a
result confirming other independent observations (Higdon priv. comm.).
If one assumes that the H$\alpha$ emission originates in normal \ion{H}{ii}
regions, could the observed LW3 emission be explained in terms of warm
dust heated by young stars?  We can estimate the amount of LW3 flux
produced by star formation in the nucleus if we use the LW3 to
H$\alpha$ correlation found in M51 (Sauvage et al.~\cite{sauvage}).
Supposing that the H$\alpha$ flux detected by Amram et
al. (\cite{amram}, see Fig.~1), is distributed over the inner ring
annulus (area $\sim$170 arcsec$^{2}$) of the Cartwheel and that
typically LW3/H$\alpha$ $\sim$ 30, we find that {\em the predicted LW3
is about 3.5 mJy} , which is of the same order as the observed LW3
emission from the Cartwheel center.  Hence, it appears that despite its
rather low-intensity, star formation activity from the Cartwheel center is
sufficient to heat the dust and  produce the observed mid-IR emission.

Another possible source of dust heating in the nuclear region could be
attributed to the infall of gas clouds. This was proposed by Struck et
al. (\cite{struck}) based on the morphology of kiloparsec size,
cometary-like structures with blue luminosities in the range
1.1--1.7$\times 10^{40}$\,ergs\,s$^{-1}$, detected in the edge of the
inner ring. An order of magnitude calculation by the authors suggests
that the dissipation of the kinetic energy of the accreting clouds via
shocks, would be $\sim$10$^{40}$\,ergs\,s$^{-1}$, sufficient to
generate a fraction of the observed blue luminosities.

\section{Conclusions} 
 
We have obtained mid-IR ISOCAM broad-band images of the Cartwheel 
group and comparing our data with our sample of normal and active 
galaxies we were able to draw the following conclusions:

1) A large segment of the outer ring is detected in the LW2 filter
which is mainly dominated by thermally-spiked PAH emission bands,
while at longer wavelengths (LW3 filter), where the emission is
primarily due to dust grains in nearly thermal equilibrium, the main
source of emission originates from a single hot-spot in the ring
associated with a particularly bright complex of \ion{H}{ii} regions.
The hot-spot has an exceptional mid-IR broad-band diagnostic ratio
LW3/LW2 of 5.2 which is among the highest of any region in the
CAMACTIV sample and is different from other regions of the ring.

2) A large fraction of the mid-IR emission is associated with the
inner ring and nucleus of the Cartwheel, in stark contrast to that
expected from optical emission-line studies (where most of the line
emission originates from the outer ring). Recently faint H$\alpha$
emission has been detected from the inner ring and it is possible that
this may be due to a low-level star formation which heats the
grains. However, in order to explain why the nuclear emission is so
powerful compared with the outer ring at ISOCAM wavelengths, it seems
that the dust of the outer ring must be spatially distributed very
differently. One possibility is that the grains in the outer ring
experience a significantly diluted UV radiation field because they are
lifted out of the disk by stronger stellar winds. Alternatively, the
nuclear regions may be heated by a very different process than the
outer, for example shock waves from infalling clouds (Struck et al
\cite{struck}).

3) The mid-IR emission from the two companions is typical for their 
Hubble type. 
 
Even though our observations shed some more light to the properties of
the hot dust in the Cartwheel galaxy, the amount and spatial
distribution of the cold dust remain uncertain. The 100\,$\mu m$ IRAS
flux of the galaxy is 1.6\,Jy but previous efforts to detect CO
emission (Horellou~\cite{horellou}) were unsuccessful, setting an
upper limit to the H$_{2}$ mass of 1.5\,10$^{9}$ M$_{\odot}$. Could
this be explained by the low metallicity of the system, by its
intrinsically low molecular gas content or simply by the large distance
of the Cartwheel? Where is the peak of the spectral energy
distribution in this galaxy? Deep sub-mm and mm wave observations
which are scheduled in the near future should enable us to address
these questions.

\begin{acknowledgements}
The authors are grateful to F. Combes (Obs. de Paris), C. Struck (Iowa
State Univ.), C. Horellou (Onsala Space Obs.), and J. Higdon (Kapteyn
Institute) for comments and stimulating discussions, as well as the
referee for useful suggestions which improved this paper. VC would
like to acknowledge the financial support from the TMR fellowship
grant ERBFMBICT960967 and the help of all members of the ISOCAM team
at CEA-Saclay.
\end{acknowledgements} 
%

\end{document}